
\vglue 1.5cm
\vskip 1cm
\centerline {The Wigner-Weyl formalism and the relativistic semi-classical
approximation \footnote1{Work supported in part by CEC Science project
SC1-CT91-0729.}}
\vskip 2cm
\centerline {J. Mourad \footnote*{E-mail:MOURAD@QCD.UPS.CIRCE.FR}}
\centerline {Laboratoire de Physique Th\'eorique et Hautes Energies
\footnote2{Laboratoire associ\'e au CNRS.}}
\centerline{Universit\'e de Paris-Sud,  b\^ at 211,
91405 Orsay, France.}
\vskip 1cm
\noindent
{\bf Abstract:}
  The relativistic semi-classical approximation for a free massive
 particle is studied
using the Wigner-Weyl formalism. A non-covariant Wigner function
is defined using the
Newton-Wigner position operator. The perturbative solution for the time
evolution is found. Causality is found to be perturbatively respected.
\vfill
\eject
\vskip .5cm

\beginsection {1 Introduction.}

The Wigner-Weyl formalism [1] offers a classical-like formulation of quantum
mechanics using phase space functions as observables and the Wigner
function as an analogue of the Liouville density function. It has been
used to study semi-classical non-relativistic quantum mechanics for time
evolution [2] and spectra determination [3]. Relativistic generalizations of
the
Wigner function have been proposed [4], they are manifestly covariant and
are not suitable to study the semi-classical approximation. The problem
lies in the fact that the natural position
operator in a manifestly covariant representation of a wave equation (as
is the Dirac representation for a spin 1/2 particle) does not have a
clear semi-classical limit; the Dirac position operator, for instance,
gives rise to a velocity that is always equal to the speed of
light. Another drawback of the manifestly covariant representation that is
related to the preceding one is that it has more than one irreducible
representation of the Poincar\'e group corresponding to more than one
particle (an electron and a positron for the Dirac case).
In this letter we propose a new definition of the Wigner function that
uses a Foldy-Wouthuysen representation [5] or equivalently the Newton-Wigner
position operator [6]. We solve pertubatively its time evolution and conclude
with some remarks.

\beginsection{2 Wigner-Weyl formalism}

The Wigner-Weyl formalism associates a phase space function to an
operator acting on Hilbert space :
$$ \hat A \rightarrow a(x,p),$$
such that
$$ \hat P \rightarrow p$$
$$ \hat X \rightarrow x,$$
$\hat P$ and $\hat X$ are respectively the momentum and position
operator. The wave function or the density matrix are replaced by  the Wigner
function:
$$ \hat \rho \rightarrow f(x,p) .$$
This mapping has the following fundamental properties:
$$ Tr(\hat A \hat \rho )=\int {dxdp \ af}  ,\eqno (2.1)$$
$$ \hat A \hat B \rightarrow ab + O(\hbar)   ,\eqno (2.2) $$
$$ {-i\over {\hbar}}[\hat A,\hat B] \rightarrow \{a,b\}+O(\hbar) .\eqno (2.3)$$
This mapping offers a way to express quantum mechanics using a classical
language [1], making clear the two conceptual differences between the
classical and the quantum theory:

-A kinematical one due to the fact
that the Wigner function and the Liouville density do not belong to the
same space, a density function does not correspond in general to a
positive density matrix via the Wigner-Weyl correspondance,  nor does a
Wigner function have to be positive. As an example the function
$$f=\delta (x) \delta (p)$$
is a Liouville density but not a Wigner function (it violates the
uncertainty relations).

- A dynamical one due to
the different equations obeyed by the the two functions. For a free
non-relativistic particle there is only a kinematical difference.

In this letter
we are only interested in the dynamical semiclassical limit, one can
take a Wigner function that is also a density function at $t=0$ and compare the
classical and quantum evolution.

 In order for the observable and its corresponding phase space
function to have the same physical content, we add the requirement
$$ \hat X _{NW} \rightarrow x ,\eqno (2.4)$$
where $\hat X _{NW}$ is the Newton-Wigner position operator. We make the
assumption that the physical position is described by the Newton-Wigner
position operator.  This is the main
difference between  our Wigner-Weyl correspondance and the manifestly
covariant one, where the manifestly covariant position operator plays
the role of $\hat X_{NW}$.
We now give the formulae relating the phase space function and the Wigner
function with the
quantum operator and the density matrix in a representation independant form:

$$ a(x,p)={2}Tr\left (\hat A e^{{2i
\over \hbar} \left (\hat
X_{NW}p-x\hat P \right )}\hat {\cal P}\right ),\eqno (2.5a)$$
$$ f(x,p)={1 \over {\left (\pi \hbar \right )}^d}Tr\left (\hat \rho e^{{2i
\over \hbar} \left (\hat
X_{NW}p-x\hat P \right )}\hat {\cal P}\right ),\eqno (2.5b)$$
d is the space dimension, $\hat P$ is the momentum operator and $\hat
{\cal P}$ is the parity operator.  These formulae (2.5) apply to a spin zero
massive particle, generalization to spinning massive particles is
straightforward.
Expressing  the Newton-Wigner operator is easy in the Foldy
representation [6], where the time
evolution of the wave function is given by (c=1)
$$i{\partial}_t \psi= (p^2+m^2)^{1\over 2}\psi ,\eqno (2.6)$$
the scalar product is:
$$\int {dp \ \psi^*\phi},$$
and the Newton-Wigner operator is:
$$ \hat X_{NW} =i{{{\partial}} \over {\partial}P}. \eqno (2.7)$$
This representation may be used to write the Wigner function in a more
familiar way, when the density matrix is a pure state, evaluation of
(2.5b) gives
$$ f(x,p)={1 \over (\pi \hbar)^d}\int {dq \psi ^{*}(p-{q \over 2})
\psi(p+{q \over 2})e^{{-i \over \hbar}qx}}.\eqno (2.8)$$
Note that this function is not manifestly covariant, it becomes so in
the classical approximation.

\beginsection {3 Time evolution}

We restrict ourselves to a one dimensional spinless massive particle.
The equation governing the time evolution of the Wigner function is
easiley seen to be
$$ {\partial}_t f_t={i\over \hbar}\left [h \left (p+{i\hbar \over
2}{{\partial}\over {\partial}x}\right ) -h \left (p-{i\hbar \over
2}{{\partial}\over {\partial}x}\right )\right ]f_t ,\eqno (3.1)$$
$$ h(p)=(p^2+m^2)^{1\over 2}.\eqno (3.2)$$
Expansion in powers of $\hbar$ gives the following equation:
$$ {\partial}_tf_t=-{p \over h(p)}{{\partial} \over {\partial }x}f_t \
-\sum_{n=1}^{\infty} {\left ({i\hbar \over 2}\right
)^{2n}}{{h^{(2n+1)}}(p)\over(2n+1)!}
{ { {\partial} ^{2n+1}}\over {{\partial}{x^{2n+1}}}}f_t, \eqno (3.3)$$
where $h^{(m)}(p)$ is the the $m^{th}$ derivative of $h(p)$.
The first term on the right hand side is the classical Liouville term
and the second gives the quantum correction to the classical motion. Note
that in the non-relativistic limit only the first term survives.
One can solve this equation perturbatively and gets the following
solution:
$$ f_{t}(x,p)=\hat Q(t)f_t^{cl}(x,p) ,\eqno (3.4)$$
$$\hat Q(t)=1+\sum_{n=1}^{\infty} \hbar^{2n}\hat Q_{n}(t),\eqno(3.5)$$
where $f_t^{cl} $ is the classical solution and $\hat Q_{n}(t)$ is a
time-dependant differential operator that we can calculate using the
following recursion formula
$$\hat Q_{1}=t{{h^{(3)}(p)} \over24}{\partial^{3}\over{\partial x^{3}}},$$
$$\hat Q_{n}={(-1)^{n+1}t \over {2^{2n}(2n+1)!}} h^{(2n+1)}(p)
{{\partial}^{2n+1}\over {\partial x^{2n+1}}}
- \sum _{p=1}^{n-1}\int_{0}^{-t}d\tau \hat Q^{ \prime} _{p}(\tau)\hat
Q_{n-p}(-\tau) \ ,\eqno (3.6)$$
where the prime is for the time derivative.
The $\hbar$ expansion giving the solution is an asymptotic one, the first
few terms give a good approximation of the exact solution  if the following
conditions are
satisfied:
$$ \Delta x \gg {\hbar \over m}=\lambda,\ \ t\ll \Delta x {\left ({\Delta x
\over \lambda}\right )}^2 ,$$
$\Delta x$ is the characteristic scale of $f_t$.
Note that to any finite order in $\hbar $ causality is respected since
the quautum solution is obtained from the classical one by application
of a differential operator; causality violation [7] is non-perturbative in
$\hbar$.
\beginsection {4 Summary and conclusion}

We defined a new non-covariant Wigner function that is suitable to
study the semiclassical limit for a relativistic particle. A representation
 independant form has been given for this function and for other phase
space functions. In the Foldy representation where positive and negative
energy states are decoupled, the expression for the Wigner function
looks
like the non-relativistic one. The equation governing its free time evolution
was  perturbatively solved. Quantum corrections to the Liouville
dynamics become important for long times or states better localized  than
the Compton wavelenght. One can consider a particle in an external
potential, whenever this potential is a function of $\hat X_{NW}$ no
complications arise, but if one considers more realistic potentials, as
the usual Coulomb potential, one has a coupling between positive and
negative energy states and  relativistic quantum mechanics of a
particle looses it validity.

\beginsection {Acknowledgements}

I would like to thank Pr R. Omnes for discussions.

\beginsection {References}

\item {[1]} H. Weyl, Z. Phys. 46 (1927) 1;
E.P Wigner, Phys. Rev. 40 (1932) 749; G.A. Baker
 Jr., Phys. Rev. 109 (1958) 2196; J.E. Moyal, Proc. Camb. Philos. Soc.
45 (1949) 99; for reviews, see N.L. Balazs and B.K.
Jennings, Phys. Rep. 104 (1984) 347; M.Hillery, R.F.
 O'Connell, M. Scully, and E.P. Wigner, Phys. Rep. 106 (1984) 121.

\item {[2]} E.J. Heller, J. Chem. Phys. 65 (1976) 1289; ibid. 67
(1977) 3339; M.V. Berry and N.L. Balazs, J. Phys. A 12  (1979) 625; R. Omnes,
 J. Stat. Phys. 57 (1989) 357.

\item {[3]} A. Voros, Annls. Inst. Poincar\'e xxiv (1976) 31; ibid.
xxvi (1977) 343; M.V. Berry, Philos. Trans. R. Soc. 287 (1977) 237.

\item {[4]} S. de Groot, La transform\'ee de Weyl et la fonction de
Wigner: une forme alternative de la m\'ecanique quantique (Les presses de
l'universit\'e de Montr\'eal, 1974) ; S. de Groot, W.A. van Leeuwen and
C.G. Van Weert, Relativistic kinetic theory (North Holland, Amsterdam,
1980);
 P.R. Holland, A. Kypriandis, Z. Maric and J.P.
Vigier, Phys. Rev. A 33 (1986) 4380; R. Hakim and H. Sivak, Ann. Phys. (N.Y)
139 (1982) 230, and refrences therein.

\item {[5]} L.L. Foldy and S.A. Wouthuysen Phys. Rev 78 (1950) 29 ; L.L. Foldy
Phys. Rev. 102 (1956) 568.

\item {[6]}T.D. Newton and E.P. Wigner, Rev. Mod. Phys. 21 (1949) 400.

\item {[7]}G.C. Hegerfeldt, Phys. Rev. D 10 (1974) 3220;
G.C. Hegerfeldt and
S.N.M. Ruijsenaars, Phys. Rev. D 22 (1980) 377;
G.C. Hegerfeldt, Phys.
Rev. Lett. 54 (1985) 2395.

\bye